# Complementary Quark and Lepton Deviation from Mass-Degeneracy Hierarchies

E. M. Lipmanov
40 Wallingford Road # 272, Brighton MA 02135, USA

**Abstract**

Near magnitudes of Dirac particle mass-ratios, mixing hierarchy-quantities and Electro-Weak charges against the background of highly successful flavor-universal one-generation EW theory is a puzzle in need of diverse inclusive research. In this paper I study the problem in proper terms of lepton and quark Deviation-from-Mass-Degeneracy (DMD) hierarchies at tree EW approximation. As primary are considered not discrete flavor symmetry but rather the deviations from mass-degeneracy-symmetry without inventing exact particular symmetry. Empirically suggested benchmark flavor patterns (zero approximation) and deviations from benchmarks caused by emergence of a small related to EW charges parameter are considered two sources of realistic flavor quantities. Physically interesting mass and mixing flavor quantities are obtained as solutions of linear and quadratic DMD-hierarchy equation-pairs with complementary patterns of quark and lepton DMD-hierarchies. Dual relations between DMD-quantities of quarks and charged leptons (Dirac particles), on the one hand, and neutrinos (likely Majorana particles), on the other hand, are inferences. Considered in the literature approximate quark-neutrino mixing angle complementarity appears naturally from the violation of benchmark patterns by the emergent small parameter.

1. **Introduction**

Large lepton and quark mass and mixing hierarchies are known problems in framework of the one-generation standard model and electroweak (EW) theory [1] extended to three generations [1_I]; they are probably far from certain theoretical explanation despite of many different symmetry-based studies[1]. The relations between flavor hierarchies and EW charges are rarely addressed; they likely need inclusive research, especially diverse ones. *Uniting* phenomenological explanation of these problems, if possible, may stimulate new theoretical solutions. An attempt in that direction at tree EW approximation [2] is based on the notion of suggested by experimental data benchmark (zero approximation) flavor pattern of elementary particles with departure from it generated by EW charges; it leads to a realistic flavor pattern in terms of DMD-quantities[2] that are useful ones in particle flavor phenomenology[3].

In this paper both lepton and quark mass and mixing hierarchy patterns are described in DMD-terms as deviated from benchmark flavor patterns by one small parameter related to EW charges. It is an extension to quarks of the flavor-electroweak lepton phenomenology [2], [3]. The extension is based on a guiding idea of exact complementary relations between quark and lepton benchmark DMD-hierarchies. More extensive motivation for the study is given in ref. [2].

---

[1]  E.g. reference [9].

[2]  The DMD-quantities are pertinent for describing mass hierarchies; 'large hierarchies' and 'order-one hierarchies' are uniquely related to large and small DMD-quantities.

[3]  In terms of DMD-quantities we deal directly with observable effects of violation of flavor symmetry without an exact form of flavor symmetry.

In Sec.3 benchmark flavor patterns of leptons and quarks with complementary DMD-hierarchies are stated. In Sec.4 ε-bound equations[4] for realistic DMD-quantities of charged leptons (CL), neutrinos and up- and down- quarks are derived, solved and compared with experimental data. In Sec.5 the main inferences are listed as flavor regularities. Conclusions are in Sec.6.

2. **Definitions of lepton and quark DMD-quantities**

The particle flavor DMD-quantities and hierarchies for three generations of CL, neutrinos and up- and down- quarks are given by definitions[5]:

$$\text{DMD(CL)}1 \equiv [(m_\tau/m_\mu)-1], \quad \text{DMD(CL)}2 \equiv [(m_\mu/m_e)-1], \quad (1)$$

$$\text{DMDH}^{(li)}(\text{CL}) \equiv \text{DMD(CL)}1 / \text{DMD(CL)}2, \quad (2)$$

$$\text{DMDH}^{(qd)}(\text{CL}) \equiv [\text{DMD(CL)}1]^2 / [\text{DMD(CL)}2]; \quad (3)$$

$$\text{DMD}(\nu)1 \equiv [(m_3^2/m_2^2)-1], \quad \text{DMD}(\nu)2 \equiv [(m_2^2/m_1^2)-1], \quad (4)$$

$$\text{DMDH}^{(li)}(\nu) \equiv \text{DMD}(\nu)2 / \text{DMD}(\nu)1, \quad (5)$$

$$\text{DMDH}^{(qd)}(\nu) \equiv [\text{DMD}(\nu)1]^2 / [\text{DMD}(\nu)2]; \quad (6)$$

$$\text{DMD(up)}1 = [(m_t/m_c)-1], \quad \text{DMD(up)}2 = [(m_c/m_u)-1], \quad (7)$$

$$\text{DMDH}^{(li)}(\text{up}) \equiv \text{DMD(up)}1 / \text{DMD(up)}2, \quad (8)$$

$$\text{DMDH}^{(qd)}(\text{up}) \equiv [\text{DMD(up)}1]^2 / [\text{DMD(up)}2]; \quad (9)$$

$$\text{DMD(dn)}1 = [(m_b/m_s)-1], \quad \text{DMD(dn)}2 = [(m_s/m_d)-1], \quad (10)$$

$$\text{DMDH}^{(li)}(\text{dn}) \equiv \text{DMD(dn)}2 / \text{DMD(dn)}1, \quad (11)$$

$$\text{DMDH}^{(qd)}(\text{dn}) \equiv [\text{DMD(dn)}2]^2 / [\text{DMD(dn)}1], \quad (12)$$

---

[4] The ε-parameter is approximately related to the low energy dynamical EW quantities - fine structure constant $\alpha \cong \varepsilon^2$ and its semi-weak analogue $\alpha_W \cong |\varepsilon^2 \log \varepsilon^2|$, e.g. [3].

[5] To remember, the DMD-quantities with number 1 always contain the largest particle mass.

where $m_e$, $m_\mu$ and $m_\tau$ are the CL masses, $m_1$, $m_2$, $m_3$ are organized three neutrino masses $m_1 < m_2 < m_3$; $m_t$, $m_c$, $m_u$ are the up-quark masses and $m_b$, $m_s$, $m_d$ are the down-quark masses; $DMDH^{(li)}(…)$ and $DMDH^{(qd)}(…)$ are the definitions of 'linear' and 'quadratic' DMD-hierarchies. For definiteness, the neutrino mass ordering (hierarchy) is chosen normal.

### 3. Complementary benchmark DMD-hierarchies of leptons and quarks

By definition the charged lepton, neutrino and up- and down-quark benchmark (zero approximation in ε) DMD-quantities and hierarchies are

1). Lepton benchmark patterns,

$$m_\nu \cong 0, \quad m_e - \text{finite}, \quad m_\mu, m_\tau \cong \infty, \tag{13}$$

$$DMD(CL)1, \quad DMD(CL)2 \cong \infty, \tag{14}$$

$$DMD(\nu)1, \quad DMD(\nu)2 \cong 0, \tag{15}$$

$$DMDH^{(qdr)}(CL; \nu) = \sqrt{2}, \quad DMDH^{(qdr)}(\theta) = 2, \tag{16}$$

$$\cos^2(2\theta_{12}) = 0, \quad \cos^2(2\theta_{23}) = 0, \quad \sin^2(2\theta_{13}) = 0, \tag{17}$$

$$\begin{pmatrix} 1/\sqrt{2} & 1/\sqrt{2} & 0 \\ -1/2 & 1/2 & 1/\sqrt{2} \\ 1/2 & -1/2 & 1/\sqrt{2} \end{pmatrix} \nu. \tag{17'}$$

Neutrino mixing DMD-quantities (17) and mixing matrix (17') are extrapolated empirically large solar $\theta_{12}$ and atmospheric $\theta_{23}$ and small reactor $\theta_{13}$ neutrino oscillation mixing angles. The relations (16) with definitions (3) and (6) refer to two lepton quadratic mass-ratio DMD-hierarchies for CL and neutrinos, and one similar quadratic DMD-hierarchy for neutrino mixing angles with definitions:

$$[DMD(\theta)1] = \cos^2(2\theta_{23}), \quad [DMD(\theta)2] = \cos^2(2\theta_{12}). \tag{18}$$

The benchmark lepton mixing matrix (17') has the known bimaximal form[6]. Bimaximal benchmark lepton mixing must be introduced here for the very definition of neutrino mixing DMD-quantities (17).

Note that from definitions (14), (15) and the quadratic benchmark hierarchies (16) follows that linear lepton DMDH-hierarchies given by definitions (2) and (5) are equal zero at benchmark pattern:

$$DMDH^{(li)}(CL) = 0, \qquad (19)$$

$$DMDH^{(li)}(\nu; \theta) = 0. \qquad (20)$$

Thus, the quadratic lepton benchmark DMD-hierarchies are of order 1 while the linear ones are infinitely large[7].

2) Quark benchmark pattern,

$$m_u, m_d - \text{finite}; \; m_q^{(up)}, m_q^{(dn)} \cong \infty, \; q > u, d, \qquad (21)$$

$$DMD(up; down)1, \; DMD(up; down)2 \cong \infty, \qquad (22)$$

$$DMDH^{(li)}(up) \cong c_q, \; DMDH^{(li)}(down) \cong c_q', \qquad (23)$$

$$\sin^2(2\theta_c) = 0, \; \sin^2(2\theta') = 0, \qquad (24)$$

$$\begin{pmatrix} 1 & 0 & 0 \\ 0 & 1 & 0 \\ 0 & 0 & 1 \end{pmatrix}_q. \qquad (24')$$

The parameters $c_q$ and $c_q'$ in (23) are constants of order 1; the notation is chosen in analogy with the definitions of CL and neutrino linear DMD-hierarchies. It suggests that $c_i$ (i = up- and down-quarks, CL and neutrinos) in (23), and (28), (32) below, are approximately close to coupling constants of the strongest interactions in the SM at energy scales defined by involved particle masses. Note that the relations of the constants $c_q$ and $c_q'$ to $\alpha_s$ have only speculative meaning.

---

[6] Bimaximal neutrino mixing was widely discussed in the literature (see e.g. [8]) as a symmetric approximate description of the large neutrino mixing. Here it is considered as pre-dynamical neutrino (probably Majorana) benchmark maximal mixing, which the deviation (caused by small ε–parameter) is counted from.

[7] Note that both large and small magnitudes of DMD-ratios have physical meaning of 'large hierarchies' in contrast to 'order 1 hierarchies'.

The quark benchmark mixing matrix (24) means no mixing of infinitely divided by mass quark benchmark mass eigenstates that is quite natural.

From the definitions (22) and finite linear quark benchmark hierarchies (23) follows that the inverse quadratic DMDH$^{(qd)}$ hierarchies defined by (9) and (12) are equal zero at benchmark level:

$$[DMD(up)2] / [DMD(up)1]^2 = 0, \qquad (25)$$

$$[DMD(dn)1] / [DMD(dn)2]^2 = 0. \qquad (26)$$

As a result, the linear quark benchmark DMD-hierarchies are of order 1 while the quadratic ones are infinitely large in contrast to the lepton case.

The benchmark DMD-hierarchies of leptons and quarks appear remarkably different. Their relations may be characterized by the term 'complementary'. These extreme benchmark quark-lepton DMD-hierarchy complementarities should remain approximately valid after deviation from benchmark flavor patterns by the small parameter $\varepsilon$.

## 4. Realistic flavor patterns of leptons and quarks

From the fact of large mass ratios of quarks and CL follows that in realistic flavor phenomenology it must be at least one large dimensionless parameter for large scale (and its reverse -for small scale). In contrast to mass ratios, description of particle mass hierarchies in terms of DMD-quantities fits well to the dual meaning of large and small scales; large DMD-quantities mean large hierarchies while small DMD-quantities mean order-one hierarchies.

The main point is that one small empirical universal dimensionless ε-parameter

$$\varepsilon \cong 0.082 \cong \exp(-5/2) \ll 1 \qquad (27)$$

transforms the extreme lepton and quark benchmark flavor patterns into finite realistic flavor patterns.

### 1. Charged lepton mass ratios

Since the benchmark value of the linear CL DMDH-hierarchy (19) is zero and taking into account data values [4] of CL masses the linear hierarchy at finite ε-parameter should be

$$\text{DMDH}^{(1i)}(\text{CL}) = \varepsilon \cong \sqrt{\alpha}, \qquad (28)$$

with α – the fine structure constant at zero momentum transfer (at the photon propagator pole value). Adding CL quadratic hierarchy (16), we get a full set of equations for realistic values of CL DMD-quantities:

$$[\text{DMD}(1)] / [\text{DMD}(2)]_{\text{CL}} \cong \varepsilon, \qquad (29)$$

$$[\text{DMD}(1)]^2 / [\text{DMD}(2)]_{\text{CL}} = \sqrt{2}. \qquad (30)$$

Solution of the set (29)-(30) for CL mass ratios is

$$m_\mu/m_e \cong \sqrt{2}/\varepsilon^2 \cong 210, \quad m_\tau/m_\mu \cong \sqrt{2}/\varepsilon \cong 17.2 \qquad (31)$$

in decent agreement with data values.

Results (31) are obtained from the idea that the realistic particle DMD-hierarchies follow from the 'benchmark' ones [2, 3] (at ε =0) after emergence of the small parameter ε ≠ 0. But this statement determines only the main approximations in (31). Obviously, they may be supplied by order one factors that approach 1 in the limit $\varepsilon \to 0$. With the latter requirement, a very simple and much more accurate choice for CL mass ratios is

$$DMD1 = (\sqrt{2}/\varepsilon)(1 - \varepsilon), \qquad (32)$$

$$DMD2 = (\sqrt{2}/\varepsilon^2)(1 - 3\varepsilon^2). \qquad (33)$$

It enhanced the accuracy[8] of tau-muon mass-ratio by three orders of magnitude, from ~0.08 to ~1 x $10^{-4}$,

$$x_1 \equiv DMD1 + 1 \cong 16.814435, \qquad (34)$$

that predicts the value of the τ-lepton mass - $m_\tau \cong 1776.59$ MeV.

The accuracy of the muon-electron mass ratio (31) is enhanced from ~0.02 to ~6 x $10^{-4}$,

$$x_2 \equiv DMD2 + 1 \cong 206.6452618. \qquad (35)$$

The relations (32) and (33) are regularities of bare CL mass-ratio quantities in flavor phenomenology at tree EW approximation, i.e. 'before' the onset of EW radiative corrections. It seems that such regularities may have physical meaning at powers of the small ε-parameter not exceeding $\varepsilon^2 \cong \alpha$ - as it is just the case in the relations (32) and (33) - since it seems improbable for the perturbative EW radiative corrections to be organized in such especially regular way.

Another remarkable regularity in CL flavor phenomenology (that is very probable of the same nature) is the accurate relation between the two CL mass ratios $x_1$ and $x_2$ (known as the Koide formula[9] [10]):

$$3[1 + x_2(1 + x_1)] = 2[1 + \sqrt{x_2}(1 + \sqrt{x_1})]^2. \qquad (36)$$

It appears that the relations for CL DMD-quantities (32) and

$$(m_e + m_\mu + m_\tau) = (2/3)(\sqrt{m_e} + \sqrt{m_\mu} + \sqrt{m_\tau})^2.$$

---

[8] It seems impossible to obtain comparable high accurate and simple enhancements in terms of particle mass ratios (not DMD-quantities).

[9] The Koide formula was originally put [10] in symmetric CL mass form:

(33) satisfy the Koide equation (36) to within a high accuracy[10] of $\sim 4 \times 10^{-5}$,

$$\{3[1+x_2(1+x_1)] - 2[1+\sqrt{x_2}(1+\sqrt{x_1})]^2\} / 3[1+x_2(1+x_1)] \cong 4 \times 10^{-5}.$$

Another point is that if instead of (35) we choose the experimental data value of the muon-electron mass-ratio $(x_2)^{exp}$ = 206.768284 (an increase in $x_2$-accuracy by four orders of magnitude) with the same tau-muon value (34), the accuracy of the relation (36) changes just a little – from $4 \times 10^{-5}$ to $3.2 \times 10^{-5}$. It means that the accuracy of Koide relation (36) is mainly determined by the tau-muon mass ratio $x_1$ (34)) and depends only weakly on the exact muon-electron mass ratio $x_2$.

Results (32) and (34) predict the τ-lepton mass

$$m_\tau \cong 1776.59 \text{ MeV}. \qquad (37)$$

## 2. Neutrino mass ratios

In case of neutrinos the extension of the linear mass-ratio DMDH-hierarchy (20) should be

$$\text{DMDH}^{(1i)}(\nu) \cong \varepsilon \sqrt{5} \cong \sqrt{\alpha_W}, \qquad (38)$$

with $\alpha_W$ – the semi-weak analog of $\alpha$ at the pole value of Z-boson propagator [3]. By combining this relation with quadratic hierarchy (16), a full set of equations for neutrino mass-ratio DMD-quantities is obtained,

$$[\text{DMD}(2)] / [\text{DMD}(1)]_\nu \cong (5\varepsilon^2) \qquad (39)$$

$$[\text{DMD}(1)]^2 / [\text{DMD}(2)]_\nu = 2. \qquad (40)$$

---

[10] Precise solution [10] of Eq.(36) for $x_1$ and $(x_2)^{exp}$ is $\cong 16..818061$ with tau mass $m_\tau \cong 1776.97$ MeV. It is about 1 S.D. from central experimental data value [4].

A special difference between the CL set (29)-(30) and neutrino one (39)-(40) is the reverse order of the terms DMD(1) and DMD(2) in the linear hierarchies. This difference is substantial and follows from detailed analysis [2] and comparison with known neutrino oscillation data [6] especially for the value of the neutrino oscillation hierarchy parameter $r = \Delta m^2_{sol}/\Delta m^2_{atm}$.

The solution of equations (39)-(40) for neutrino DMD-quantities is given by

$$DMD(\nu)1 \equiv [(m_3^2/m_2^2)-1] \cong 2(5\varepsilon^2) \cong 0.067 \ll 1, \qquad (41)$$

$$DMD(\nu)2 \equiv [(m_2^2/m_1^2)-1] \cong 2(5\varepsilon^2)^2 \cong 0.0023 \ll 1. \qquad (42)$$

The main result from these solutions is that the neutrino masses are quasi-degenerate

$$m_2^2/m_1^2 \cong 1, \quad m_3^2/m_2^2 \cong 1. \qquad (43)$$

The second result is for the magnitude of the neutrino oscillation hierarchy parameter

$$r = \Delta m^2_{sol}/\Delta m^2_{atm} \equiv DMDH^{(li)}(\nu) \cong 5\varepsilon^2 \cong 1/30 \qquad (44)$$

in reasonable agreement with best fit value from neutrino oscillation data analysis [6].

From definition (13) follows the relation $m_\nu/m_e \cong 0$ at benchmark pattern; after the deviation from benchmark by emergence of small $\varepsilon$-parameter QD-neutrinos appear with small nonzero mass scale

$$m_\nu \cong \pi \varepsilon^6 m_e / 3 \cong 0.16\,eV. \qquad (45)$$

The factor 3 in the denominator is related to neutrino quasi-degeneracy [3].

From the solutions (41)-(42) and relations between neutrino oscillation mass-squared differences and absolute neutrino masses

$$\Delta m^2_{sol} \equiv 0.0023\,m_1^2, \quad \Delta m^2_{atm} \equiv 0.067\,m_2^2, \qquad (46)$$

and neutrino mass-scale (45), quantitative estimations for neutrino oscillation mass-squared differences follow

$$\Delta m^2_{sol} \equiv 5.9 \times 10^{-5} \text{ eV}^2, \quad \Delta m^2_{atm} \equiv 1.7 \times 10^{-3} \text{ eV}^2 \qquad (47)$$

in reasonable agreement with oscillation data [6].

### 3. Quark mass ratios

Since at benchmark pattern the inverse quadratic DMDH-hierarchies (25) and (26) are zero, the realistic ones should be determined by the small parameters $(\varepsilon^2)$ for up-quarks and $(5\varepsilon^2)$ for down-quarks (the former appears in CL case whereas the latter – in the neutrino one).

A set of equations with one small parameter and one order-one parameters for realistic values of up- and down-quark DMD-quantities is given by

$$\text{DMDH}^{(li)}(\text{up}) = c_q, \qquad (48)$$

$$\text{DMDH}^{(qd)}(\text{up}) \cong c_q (1/\varepsilon^2) \cong c_q (1/\alpha), \qquad (49)$$

$$\text{DMDH}^{(li)}(\text{dn}) = c_q', \qquad (50)$$

$$\text{DMDH}^{(qd)}(\text{dn}) \cong c_q' (1/5\varepsilon^2) \cong c_q' (1/\alpha_W). \qquad (51)$$

The point is that the two parameters $(c_q, c_q')$ are supposed to be order-one parameters – in sharp contrast to the truly small universal parameter $\varepsilon$. These two order-one parameters are probably related to the strong quark interactions.

#### i) Up-quarks

Relations (48) and (49) with definitions (7) suggest a pair of equations for up-quark DMD-quantities,

$$[(DMD1)/(DMD2)]_{up} = c_q, \qquad (52)$$

$$[(DMD1)^2/(DMD2)]_{up} \cong c_q(1/\varepsilon^2), \qquad (53)$$

with solution

$$m_t/m_c \cong 1/\varepsilon^2, \quad m_c/m_u \cong 1/(\varepsilon^2 c_q) > m_t/m_c. \qquad (54)$$

At a simple example with $c_q \cong 1/3$ and $m_t \cong 172$ GeV, the masses of charm- and up- quarks are

$$m_c \approx 1.16 \text{ GeV}, \quad m_u \approx 2.6 \text{ MeV}, \quad (55)$$

compatible with data [4].

Two relations (52)-(53) for up-quark and (29)-(30) for CL DMD-hierarchies lead to interesting inferences of large hierarchies between up-quark and CL DMD-hierarchies

$$\text{DMDH}^{(li)}(up)/\text{DMDH}^{(li)}(CL) \cong (c_q/\varepsilon) \gg 1, \quad (56)$$

$$\text{DMDH}^{(qd)}(up)/\text{DMDH}^{(qd)}(CL) \cong (c_q/2)(1/\varepsilon^2) \gg 1. \quad (57)$$

### ii) **Down-quarks**

Equations (50) and (51) with definitions (10) lead to the pair of equations for down-quark DMD-quantities

$$[(DMD2)/(DMD1)]|dn = c_q', \quad (58)$$

$$[(DMD2)^2/(DMD1)]dn \cong c_q'(1/5\varepsilon^2), \quad (59)$$

with solution

$$m_b/m_s \cong 1/5\varepsilon^2 c_q', \quad m_s/m_d \cong 1/5\varepsilon^2 < m_b/m_s. \quad (60)$$

At a simple example with $a_q' \cong 0.9$ and $m_b \cong 4.2$ GeV, the masses of strange- and d-quarks are

$$m_s \cong 127.35 \text{ MeV}, \quad m_d \cong 4.29 \text{ MeV} \quad (61)$$

compatible with data [4].

Relations (58)-(59) for down-quark DMD-hierarchies and (39)-(40) for neutrino DMD-hierarchies lead to the inference of large hierarchies between down-quark and neutrino DMD-hierarchies

$$\text{DMDH}^{(li)}(dn)/\text{DMDH}^{(li)}(\nu) \cong c_q'/5\varepsilon^2 \gg 1, \quad (62)$$

$$\text{DMDH}^{(qd)}(dn)/\text{DMDH}^{(qd)}(\nu) \cong (1/5\varepsilon^2)(c_q'/2) \gg 1. \quad (63)$$

## 4. Neutrino and quark mixing patterns

### i) Neutrino mixing

At benchmark pattern the neutrino mixing angles are maximal ($\pi/4$) and obey the quadratic (16) and linear (20) DMD-hierarchy rules with definitions

$$\text{DMD 1} = \cos^2(2\theta_{12}), \quad \text{DMD 2} = \cos^2(2\theta_{23}). \quad (64)$$

With due regard of oscillation data, after emergence of the small parameter $\varepsilon$ we get a set of equations for two mixing DMD-quantities,

$$[\text{DMD 2}]/[\text{DMD 1}] \cong \varepsilon/2 \cong \sqrt{\alpha}/2, \quad (65)$$

$$[\text{DMD 1}]^2 / [\text{DMD 2}] = 4. \quad (66)$$

The different order of the terms DMD1 and DMD2 in (65) and (66) is in analogy with neutrino mass-ratio DMD set (39)-(40).

The solution of equations (64)-(66) for neutrino mixing angles are given by

$$\cos^2(2\theta_{12}) \cong 2\varepsilon, \quad \cos^2(2\theta_{23}) \cong \varepsilon^2. \quad (67)$$

Thus, quasi-degenerate neutrinos have large but not maximal solar and atmospheric mixing angles with strongly hierarchical deviations from maximal mixing in quantitative accord with oscillation data.

The solutions (67) determine the neutrino mixing matrix in terms of one parameter $\varepsilon$; in the standard representation [4] it is approximately given by

$$V_\ell \cong \begin{pmatrix} 0.838 & 0.545 & 0 \\ -0.401 & 0.616 & 0.678 \\ 0.369 & -0.568 & 0.736 \end{pmatrix} \nu. \quad (68)$$

This neutrino one-parameter mixing matrix is in good agreement with the data indications; the deviations of solar and atmospheric mixing angles from best-fit experimental values [6],

$$(\sin^2 \theta_{23})_{exp} = 0.466 + 0.178 - 0.135,$$

$$(\sin^2 \theta_{12})_{exp} = 0.312 + 0.063 - 0.049,$$

are ~1% for $\theta_{23}$ angle and ~5% for the solar one $\theta_{12}$.

ii) **Quark mixing**

The realistic quark mixing should be considered as small deviated from the benchmark minimal (zero) mixing (24); so, quark mixing DMD-quantities are given by

$$\text{DMD 1} = \sin^2(2\theta_c), \quad \text{DMD 2} = \sin^2(2\theta'). \tag{69}$$

The equations for quark mixing should be the same as for leptons (65)-(66) after the replacement of DMD-quantities as indicated in (69):

$$[\sin^2 2\theta']/[\sin^2 2\theta_c] \cong \varepsilon/2, \tag{70}$$

$$[\sin^2 2\theta_c]^2 / [\sin^2 2\theta'] = 4. \tag{71}$$

Solutions of equation set (70)-(71) are

$$\sin^2(2\theta_c) \cong 2\varepsilon, \quad \sin^2(2\theta') \cong \varepsilon^2. \tag{72}$$

These solutions determine quark mixing matrix $V_q$ through one small universal parameter $\varepsilon$:

$$V_q \cong \begin{pmatrix} 0.98 & 0.21 & 0 \\ -0.21 & 0.98 & 0.04 \\ 0.01 & -0.04 & 1 \end{pmatrix}_q. \tag{73}$$

It is in reasonable qualitative agreement with the CKM matrix from world data analysis [4]. The main disagreement is for the Cabibbo mixing in $V_{12}$ and $V_{11}$.

Compare the solutions for neutrino (67) and quark (72) mixing angles. A strait inference from that comparison is the statement of quark-neutrino mixing angle approximate complementarity [7]:

$$2\theta_{12} \cong (\pi/2 - 2\theta_c), \quad 2\theta_{23} \cong (\pi/2 - 2\theta'). \tag{74}$$

It is a result of two physical statements – exact quark-neutrino mixing angle complementarity in the benchmark (background) patterns and deviations from those patterns by the small parameter $\varepsilon$.

It should be noted that since neutrino mixing is determined in Eq.(65) by the dynamical constant $\varepsilon^2 \cong \alpha(q^2=0)$, and not $5\varepsilon^2 \cong \alpha_W(q^2=M_Z^2)$, the very deviation from maximal neutrino mixing, as mentioned above, is probably more related to neutrino SU(2)-partners. Then, the mixing matrix (73) can be also appropriate for bare CL mixing. The quark mixing matrix should be primarily equal to the CL if the decisive here feature is Dirac-Majorana relation, not the quark-lepton one.

## 5. **Inferences pointing to new flavor physics**

Main result of the present study is that linear and quadratic DMD-hierarchy equations with only one small parameter $\varepsilon$ determine the complete system of 12 lepton and quark flavor DMD-quantities, which reasonably fit to experimental data.

These equations together with hints from experimental data are defining flavor physics regularities:

1) The primary quantities in flavor phenomenology are DMD-hierarchies. Realistic linear and quadratic DMD-hierarchies are defined by empirically suggested benchmark patterns and one small universal parameter $\varepsilon$ that generates violation from benchmark; they determine all interesting DMD-quantities of neutrinos, CL and quarks[11].

---

[11] CP-violation effects are not considered here.

2) The pairs of main DMD-hierarchy equations are similar for all four types of elementary particles (ν, CL, up- and down-quarks). Linear DMD-hierarchies are always much smaller than the quadratic ones.

3) There are large hierarchies between quark and lepton DMD-hierarchies (both linear and quadratic). This large quark-lepton hierarchy of DMD-hierarchies is a characteristic difference between quarks and leptons in the empirical flavor mass-ratio phenomenology.

In contrast to the mass-ratio DMD-hierarchies, the mixing angle DMD-hierarchies of leptons and quarks are formally equal; it may be since they are related not to quark-lepton feature of elementary particles, but rather to the Dirac-Majorana one.

4) Linear DMD-hierarchies are approximately close to the coupling constants of the strongest Standard Model interactions for involved particles at energy scales related to particle masses: CL – to $\alpha$, neutrinos – to $\alpha_W$, both up- and down-quarks – probably to the strong interaction constant $\alpha_s$ at different energy scales.

5) The universal parameter $\varepsilon$ introduces two connected dimensionless scales for particle DMD-quantities: i) a small one ($\varepsilon$) for neutrino DMD-quantities and particle mixing and ii) a large one ($1/\varepsilon$) for Dirac particle mass ratios. In addition, the hierarchy of CL mass ratios is large whereas the up- and down-quark mass-ratio hierarchies are remarkably smaller. These features fit well to the empirical mass spectra of known Dirac elementary particles.

6) On the one hand, there is a particular similarity between up-quark and CL DMD-quantities for mass-ratios - both are approximately defined by $1/\varepsilon \cong 1/\sqrt{\alpha}$; on the other hand, there is a particular similarity between down-quark and neutrino mass-ratio DMD-quantities (not mass-ratios) - both are approximately defined by $1/\sqrt{(5\varepsilon^2)} \cong 1/\sqrt{\alpha_W}$.

7) Dirac-Majorana DMD-duality. It stated that Dirac particle mass-ratio-DMD-quantities are determined by the large scale ($1/\varepsilon$) whereas the Majorana neutrino ones – by the small scale ($\varepsilon$). Interestingly, it is described above by the comparatively reverse order of the DMD1 and DMD2 terms in the linear and quadratic-hierarchy relations, see (29)-(30), (52)-(53), (58)-(59)) for Dirac particles, on the one hand, and (39)-(40) for neutrinos, on the other hand. This Dirac-Majorana duality predicts QD-neutrinos and is supported by direct inferences for magnitudes of neutrino physical quantities, - 1) oscillation hierarchy parameter *r*, 2) solar and atmospheric mass-squared differences, 3) neutrino mass scale, 4) solar and atmospheric neutrino oscillation mixing angles $\theta_{12}$ and $\theta_{23}$ and 5) complementarity between the Dirac particle mixing angles and the neutrino mixing ones – they are in reasonable agreement with data indications.

In extreme form these regularities are already present at the benchmark particle flavor patterns. At $\varepsilon = 0$ the mass-ratio DMD-quantities of all Dirac particles are infinitely large whereas the ones of neutrinos are equal zero. The magnitudes of mixing angle DMD-quantities of the neutrinos and quarks at $\varepsilon = 0$ are maximally opposite. The emergence of a small, not zero, $\varepsilon$-parameter transfers these features of the benchmark flavor patterns to the realistic particle flavor patterns.

The above list of flavor regularities points to new flavor physics beyond the one-generation Standard Model with two main characteristic features i) substantial relations between SM charges and elementary particle DMD-hierarchies, and ii) Dirac-Majorana DMD-dualities.

## 6. Conclusions

The DMD-hierarchies are phenomenological means for revealing new physical effects (regularities) related to particle flavor symmetry-violations without explicit reference to particular exact discrete symmetry. In the realistic case with three flavors, linear and quadratic DMD-hierarchy equations derive all 6 basic DMD-quantity pairs (DMD1 and DMD2) for elementary particles – neutrino, CL, up- and down-quark mass ratios and neutrino and quark mixing angles.

The relation of the present DMD-phenomenology to the concept of symmetry is opposite to the well known regular one. We start with maximal symmetry violation and arrive at realistic approximate flavor physics with finite symmetry violation. It is an interesting (new physics) deviation from symmetry paradigm that is based on a new idea of a small universal flavor-electroweak parameter $\varepsilon$ and its emergence from benchmark pattern (at $\varepsilon=0$).

Since the DMD-hierarchies are determined by one small universal parameter $\varepsilon \ll 1$, there are two and only two types of possible solutions with 1) large DMD-quantities (large mass-ratios), determined by the large scale $1/\varepsilon$, that are appropriate for all known Dirac particle mass ratios and 2) small DMD-quantities (order-one mass-ratios), determined by the small scale $\varepsilon$. Empirically large mass ratios of all known Dirac particles suggest that only the neutrinos may be described by the second type solution – to be quasi-degenerate and likely of Majorana nature.

If experimental confirmation of QD-neutrino masses be made, the present 3-flavor DMD-phenomenology that smoothly incorporates QD-neutrinos in the elementary particle mass patterns may become a relevant pre-theoretical ground for new flavor physics.

As shown above, there are unity and differences between the patterns of quark and lepton DMD-hierarchies. The unity is that flavor quantities (describing mass distributions of particle-copies) of leptons and quarks are governed by similar form of linear and quadratic DMD-hierarchies. The difference is that the up- and down- quark linear and quadratic mass-ratio DMD-hierarchies are much larger than the corresponding charged lepton and neutrino ones.

Emphasized flavor regularities are summarized in Sec.5. Reasonable agreement with experimental data of the complete large system of lepton and quark DMD-quantities suggests that the studied here at tree EW approximation quadratic and linear DMD-hierarchies are not crucially destroyed by the EW and strong radiative corrections.

In the considered phenomenology neutrinos are different from CL and quarks. This difference is described by the mentioned above second type of solutions. It predicts QD-neutrino masses. But the basic hierarchy equations for neutrinos and CL are much similar with the only formal difference (that lead to all the important consequences) being the opposite relative (DMD1-DMD 2)-ordering in the linear and quadratic neutrino DMD-hierarchies. The encouraging point is that these distinct similarity and difference lead to experimentally verifiable inferences for the complete set of 3-flavor neutrino quantities such as QD-neutrino type with natural values of oscillation hierarchy parameter, small absolute neutrino mass scale and large neutrino mixing as compared with quark one.

21W. Rodejohann, Nucl. Phys. B687, 31 (2004).

[8] G. Altarelli, F. Feruglio, L. Merlo, arXiv:0903.1940.
B530, 167 (2002).

[9] C. H. Albright, arXiv:0905.0146.

[10] Y. Koide, Lett. Nuovo Cimento, 34, 201 (1982).